COCCIA **LAB**

*To discover the causes of social, economic and technological change*

# The laws of the evolution of research fields

Mario COCCIA

CNR -- NATIONAL RESEARCH COUNCIL OF ITALY

&

ARIZONA STATE UNIVERSITY

# The laws of the evolution of research fields

Coccia Mario [1, 2]

1. CNR -- National Research Council of Italy
2. Arizona State University

Current Address: CocciaLAB at CNR -- National Research Council of Italy
Collegio Carlo Alberto, Via Real Collegio, n. 30, 10024-Moncalieri (Torino), Italy

Mario Coccia ORCiD: http://orcid.org/0000-0003-1957-6731

\* E-mail: mario.coccia@cnr.it

**Abstract**. A fundamental question in the field of social studies of science is how research fields emerge, grow and decline over time and space. This study confronts this question here by developing an inductive analysis of emerging research fields represented by human microbiome, evolutionary robotics and astrobiology. In particular, number of papers from starting years to 2017 of each emerging research field is analyzed considering the subject areas (i.e., disciplines) of authors. Findings suggest some empirical laws of the evolution of research fields: the *first law* states that the evolution of a specific research field is driven by few scientific disciplines (3-5) that generate more than 80% of documents (concentration of the scientific production); the *second law* states that the evolution of research fields is path-dependent of a critical discipline (it can be a native discipline that has originated the research field or a new discipline emerged during the social dynamics of science); the *third law* states that a research field can be driven during its evolution by a new discipline originated by a process of specialization within science. The findings here can explain and generalize, whenever possible some properties of the evolution of scientific fields that are due to interaction between disciplines, convergence between basic and applied research fields and interdisciplinary in scientific research. Overall, then, this study begins the process of clarifying and generalizing, as far as possible, the properties of the social construction and evolution of science to lay a foundation for the development of sophisticated theories.

ACKNOWLEDGMENTS. I gratefully acknowledge financial support from the CNR - National Research Council of Italy for my visiting at Arizona State University (Grant CNR - NEH Memorandum n. 0072373-2014 and n. 0003005-2016) where this research started in 2016. I also thank the fruitful suggestions and comments of Ken Aiello, Sara I. Walker and seminar participants at the Beyond-Center for Fundamental Concepts in Science (Arizona State University in Tempe, USA).

**Introduction**

The purpose of this study is to analyses how research fields evolve in order to explain, whenever possible, some properties of the social construction and evolution of science.

The role of science in society has been exploded in the last fifty years (Coccia, 2018; Freeman, 1960; Merton, 1968; Ecklund et al., 2016; Stephan, 1996). Freedman (1960, p. 3) argues that: "Science is a form of human activity through pursuit of which mankind acquires an increasingly fuller and more accurate knowledge and understanding of nature, past, present and future, and an increasing capacity to adapt itself to and to change its environment and to modify its own characteristics". Kuhn (1962, pp.1-2) claimed that: "If science is the constellation of facts, theories, and methods collected in current texts, then scientists are the men who, successfully or not, have striven to contribute one or another element to that particular constellation. Scientific development becomes the piecemeal process by which these items have been added, singly and in combination, to the ever growing stockpile that constitutes scientific technique and knowledge". Kuhn (1962) also states that vital discoveries (or breakthroughs), based on a long-run development of "normal science", include not only radical changes that have a significant impact on several research fields, but also scientific changes whose consequences are within a specific scientific discipline in which the change has taken place (cf., Andersen, 1998).

Lakatos (1978, p. 168, original Italics and emphasis) argues that:

> science . . . can be regarded as a huge research program . . . .progressive and degenerating problem-shifts in series of successive theories. But in history of science we find a continuity which connects such series. . . . The programme consists of methodological rules: some tell us what paths of research to avoid (*negative heuristic*), and others what paths to pursue (*positive heuristic*) - By 'path of research' I mean an objective concept describing something in the Platonic 'third world' of ideas: a series of successive theories, each one 'eliminating' its predecessors (in footnote 57) - . . . . What I have primarily in mind is not science as a whole, but rather particular research-programmes, such as the one known as 'Cartesian metaphysics. . . . a 'metaphysical' research-programme to look behind all phenomena (and theories) for explanations based on clockwork mechanisms (positive heuristic). . . A research-programme is successful if in the process it leads to a progressive problem-shift; unsuccessful if it leads to a degenerating problem-shift . . . . Newton's gravitational theory was possibly the most successful research-programme ever (p. 169). . . . The reconstruction of scientific progress as proliferation of rival research-programmes and progressive and degenerative problem-shifts gives a picture of the scientific enterprise which is in many ways different from the picture provided by its reconstruction as a succession of bold theories and their dramatic overthrows (p. 182).

A discipline is a concept that derives from Latin *disciplina,* derivation of *discĕre*= to learn. In science, discipline is a system of norms, theories and principles, organized, systematized and based on specific methods of inquiry that investigate phenomena in nature and society. A research field is a sub-set of a discipline that investigates specific topics and/or phenomena to solve theoretical and practical problems that generate discoveries and/or science advances of applied and/or basic sciences. The complex factors supporting research fields emerge in specific contexts and subjects, such as leading universities and/or research labs, outstanding scholars and/or fruitful collaborations between scholars of different disciplines (cf., Coccia and Wang, 2016; Stephan, 1996). Many studies of science analyze international collaboration between organizations for its impetus in advancing scientific production and fostering breakthroughs. Coccia and Wang (2016, p. 2057) reveal that patterns of international collaboration are generating a convergence between applied and basic sciences.

Social studies of science argue that the origin and evolution of new scientific fields can be due to discoveries and/or scientific breakthroughs driven by interdisciplinary between applied and theoretical disciplines, such as nanomedicine (originated from interaction between physics, engineering, biology and medicine), biotechnology, etc. (Sun et al., 2013; cf., De Solla Price, 1986; Latour, 1987; Latour and Woolgar, 1979; Mulkay, 1975). Morillo et al. (2003, p. 1237) claim that science is increasing the interdisciplinary between research fields because of a higher specialization in applied/theoretical sciences and combination of bodies of knowledge directed to solve more and more complex scientific problems in nature and society. Interdisciplinary generates new social community of scholars and this is one of the vital drivers of new research fields underlying the evolution of science (Gibbons et al., 1994; Guimera et al., 2005; Klein, 1996; Sun et al., 2013; Wagner, 2008). Sun et al. (2013) state that some theories consider the social interaction among groups of scientists "as the driving force behind the evolution of disciplines" (cf., Wuchty et al., 2007). As a matter of fact, the interaction between scholars of different disciplines might be one of the factors of social construction of science to support the evolution of new research fields. In fact, Small (1999, p. 812) argues that: "crossover fields are frequently encountered, and the location of a field can occasionally defy its disciplinary origins." Moreover, in social studies of science, many scholars have suggested a hierarchy of sciences and investigated their evolution, but these studies -recently performed with comparative maps of science that represent how research fields are related and how relationships between scientific fields evolve- have also shortcomings (Boyack, 2004; Boyack et., 2005; Fanelli and Glänzel, 2013; Simonton, 2002; Small, 1999; Smith et al., 2000). In general, scientific fields are not static entities but they change dynamically over time within the social dynamics of science. Some of these changes are progressive because of the essential nature of scientific progress in society (Simonton, 2004, p. 65). Sun et al. (2013, p. 4) suggest that the evolution of science is a process driven only by social dynamics. In particular, results suggest that sociocognitive interactions of scientists and scientific communities play a vital role in shaping the origin and evolution of scientific fields (Sun et al., 2013).

Hence, the analysis of scientific fields during their evolution has a vital role to explain properties of the construction and social dynamics of science and to predict how applied and basic sciences evolve to solve problems, improve technology and satisfy human needs and desires in society (Coccia and Bellitto, 2018; Coccia and Wang, 2016; Sun et al., 2013).

In light of the continuing importance of the studies of social dynamics of science, this paper focuses specifically on following research questions:

- How do research fields evolve?

- What drives the evolution of research fields? One or more disciplines? The same discipline or different disciplines over time?

The underlying problem of these research questions is to investigate the properties of social construction and evolution of science. These topics are basic in the social studies of science because the understanding of the origin and evolution of scientific fields can guide funding for research to support science advances, new technology and human progress in society (Coccia and Bellitto, 2018; Coccia, 2018; Kitcher, 2001; Romer, 1994; Storer, 1967; Schmookler, 1966; Stephan and Levin, 1992; Sun et al. 2013)[1].

This inductive study here investigates three critical research fields: Human Microbiome (in short Microbiome), Evolutionary Robotics and Astrobiology. Results of this study may explain and generalize, whenever possible properties of the evolution of research fields to design a research policy directed to support emerging research fields and science advances for technological and economic change in society. In order to perform the study here, next section presents materials and methods.

---

[1] For other studies about the social dynamics of science and technological innovations see also, Calabrese et al., 2002, 2005; Calcaltelli et al., 2003; Cavallo et al., 2014, 2014a, 2015; Chagpar and Coccia, 2012; Coccia, 2001, 2002, 2003, 2004, 2004a, 2005, 2005a, 2005b, 2005c, 2005d, 2005e, 2005f, 2005g, 2006, 2006a, 2007, 2008, 2008a, 2009, 2009a, 2009b, 2009c, 2010, 2010a, 2010b, 2010c, 2010d, 2011, 2012, 2012a, 2012b, 2012c, 2012d, 2013, 2013a, 2014, 2014a, 2014b, 2014c, 2014d, 2014e, 2014f, 2015, 2015 a, 2015b, 2015c, 2016, 2016a, 2016b, 2017, 2017a, 2017b, 2017c, 2017d, 2017e, 2017f, 2017g, 2017h, 2017i, 2017l, 2017m, 2 018, 2018a, 2018b, 2018c, 2018d, 2018e; Coccia and Bellitto, 2018; Coccia and Bozeman, 2016; Coccia and Cadario, 2014; Coccia and Finardi, 2012; Coccia et al., 2010, 2012, 2015; Coccia and Rolfo, 2007, 2008, 2009, 2010, 2013; Coccia and Wang, 2015, 2016; Rolfo and Coccia, 2005.

**Materials and Methods**

The data of this study are from Scopus (2018). In particular, this study applies the tool by Scopus (2018) "document search" in Article title, Abstract, Keywords.

The scientific fields under study here are: microbiome, evolutionary robotics and astrobiology/exobiology. This study assumes that disciplines in science are the subject areas indicated by Scopus (2018).

The steps performed to gather data on the origin and evolution of research fields are:

1. The search in Scopus (2018) is focused on following keywords of research fields under study: "microbiome", "evolutionary robotics" and "astrobiology/exobiology" (Fig. 1).

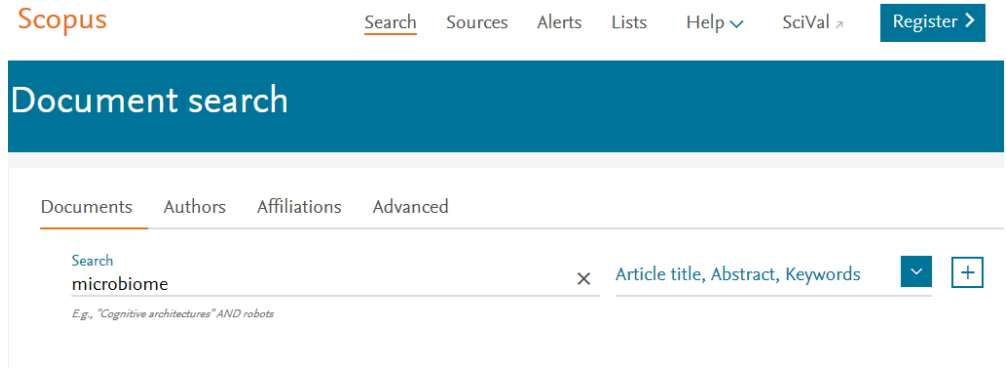

Figure 1. Document search in Scopus (2018): example of microbiome.

2. After that, the study design considers data from the first year indicated in Scopus to 2017 (year 2018 is not included here because data are in progress). Figure 2 shows the years and number of documents for the research field of "microbiome" (Scopus, 2018). Similar data are from other research fields under study.

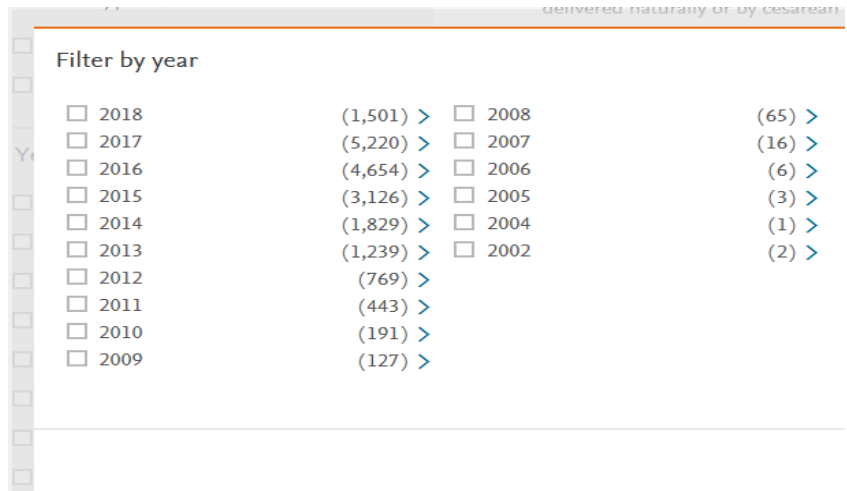

Figure 2. Years and number of documents for the research field of "microbiome"

Every year in Scopus (2018) indicates the total number of articles concerning the research field under study. The assumption here is that the year in which it is appeared the document with the first use and/or analysis of new concepts, it indicates the origin of the research field. The number of articles in subsequent years shows the evolution of research field (Fig. 2).

3. Every year in Scopus (2018) is selected and the following information are considered:

- Number of documents

- Subject areas (a proxy of disciplines underlying research fields), such as medicine, immunology and microbiology, etc. Each subject area is associated with the department/s of authors (see Figure 3).

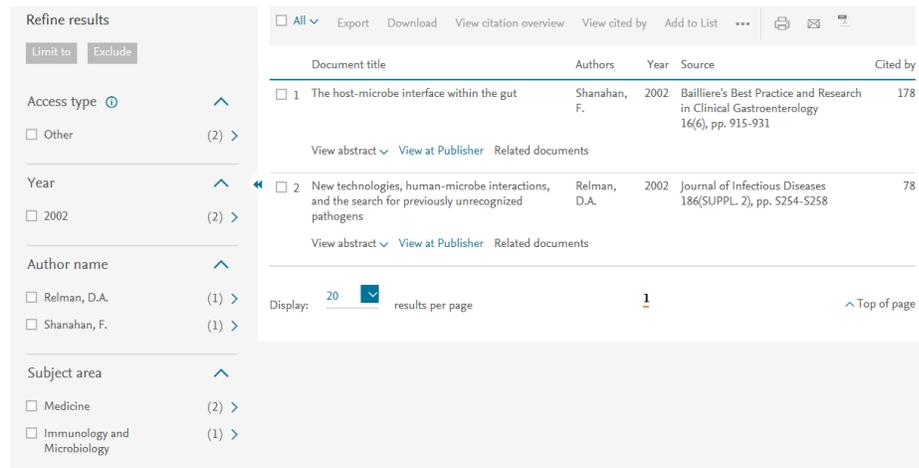

Figure 3. Information per year: example of the 2002 for the research field of microbiome

For the first year is considered the following information to detect the origin of the research field: subject areas, authors, documents, journals, departments and universities.

4. In a spreadsheet Excel, the information by Scopus (2018) are systematized as follows: the first line indicates the years, for instance from 2002 to 2017; the first column indicates the subject areas, such as medicine, biochemistry, genetics and molecular biology, etc. associated with papers of research fields under study. Each cell of the matrix indicates a number that means in a specific year the number of subject areas to which articles concerning a research field are allocated by Scopus (2018), depending on affiliations of authors. After that, these data are represented in a graph indicating the trend per year of the research field, considering the articles in subject areas supporting the research field under study ($x$-axis is time-years, $y$-axis is the total number of publications of the research field). Another graph indicates the annual weight of the occurrences of each subject area, derived from articles supporting the evolution of research field under study.

The ratio, represented on $y$-axis[2], is given by:

$$weight\ of\ occurrences\ of\ the\ subject\ area\ in\ the\ research\ field\ i\ at\ t\ =$$

$$\frac{occurrence\ of\ the\ subject\ area\ in\ the\ research\ field\ i\ at\ time\ t}{total\ of\ number\ of\ occurrences\ in\ research\ field\ at\ t} \qquad [1]$$

The study here also estimates the trends of research fields under study applying the best curve estimation model with regression analysis. The model applied can be:

- Lineal model: $y = a + bt$, series of data are modelled as a linear function of time $\qquad$ [2]

- Model of growth with equation $y = e^{a+bt}$ when data take off at a specific year $\qquad$ [3]

- Exponential model with equation $y = a\,e^{bt}$ is similar to previous model (3) $\qquad$ [4]

Regression analyses show the coefficient of determination $R^2$ and $F$ that indicates the ratio of the variance explained by the model to the unexplained variance. Models are estimated with Ordinary Least Squares (OLS). Statistical analyses are performed by using the Software IBM SPSS® Statistics 21.

Overall, then, scientific outputs from statistical analyses can provide vital information on how research fields evolve and which disciplines contribute to their evolutionary dynamics.

## Results

The scientific fields under study are: human microbiome (in short, microbiome), evolutionary robotics and astrobiology/exobiology. Empirical results can provide findings to explain the social construction and evolution of research fields and generalize, whenever possible, properties.

---

[2] $x$-axis indicates the time (years)

## 1.1 Human Microbiome

The American Microbiome Institute (2015) claims that:

> The human microbiome refers to the assemblage of microbes that live in the human body. While these microbes inhabit all parts of our body that are exposed to the environment, such as the skin, mouth, and vagina, most reside in the gut where they have a constant supply of nutrients. . . . scientists are now uncovering the significant role they play in human health. Nearly every scientific study performed that has attempted to correlate the microbiome with specific traits or diseases has been successful. In other words studies are finding that our bacteria (or lack thereof) can be linked to or associated with: obesity, malnutrition, heart disease, diabetes, celiac disease, eczema, asthma, multiple sclerosis, colitis, some cancers, and even autism.

In 2008, Human Microbiome Project is created with public research funds from US National Institutes of Health of the Department of Health and Human Services. The goal of this project is to explain the role played by human microbiome in human health and diseases and its comprehensive characterization.

The search document of the keyword "microbiome" in Scopus (2018) shows that the first articles concerning this topic is in 2002 by American microbiologist Prof. Relman David Arnold from Department of Microbiology (Stanford University, CA, United States) and US Dept. of Veterans Affairs at Palo Alto Hlth. Care Syst. in California. The article is: Relman D. A. (2002) "New technologies, human-microbe interactions, and the search for previously unrecognized pathogens", Journal of Infectious Diseases, volume 186, issue Suppl. 2, pages S254-S258

Another article in 2002 was published by Shanahan Fergus from Department of Medicine, University College Cork, National University of Ireland. Shanahan's article is: "The host-microbe interface within the gut",

Bailliere's Best Practice and Research in Clinical Gastroenterology, vol. 16, issue 6, December 2002, pages 915-931.

These articles can be considered the origin of this new research field in 2002 that is growing with geometric rates as shown in figure 4.

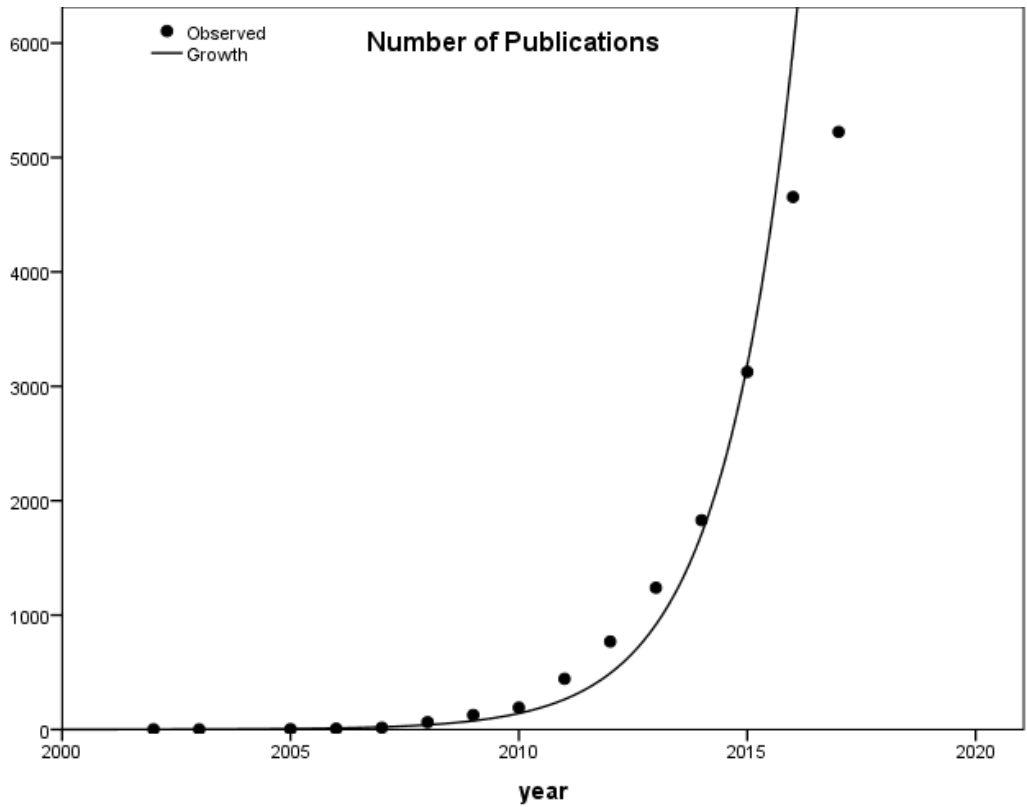

Figure 4. Growth of the research field of human microbiome

The estimated relationship with regression analysis applying a model of growth with equation [3] $y = e^{a+b\,t}$ or $ln\,y = a + bt$, is represented in table 1. The $R^2$ value and $t$ ratio are very high and model [3] explains more than 98% variance in the data.

**Table 1**: Estimated relationship of the evolution of the research field "human microbiome"

| Growth model | Constant a (St. Err.) | Coefficient b (St. Err.) | Stand. coefficient | $R^2$ | F (Sign.) |
|---|---|---|---|---|---|
| Dependent variable (D): Annually Publications | | | | | |
| *Explanatory variable* ○ Time 2002-2017 | −1246.91*** (61.35) | 0.62*** (0.031) | 0.985 | 0.97 | 416.28 (0.001) |

*Note*: ***=Coefficient is significant at *p-value<0.001*

Figure 5 shows the trend of human microbiome based on disciplines supporting its evolutionary growth. In particular, the analysis shows that medicine, biochemistry, genetics and molecular biology are the most important disciplines supporting the evolution of human microbiome.

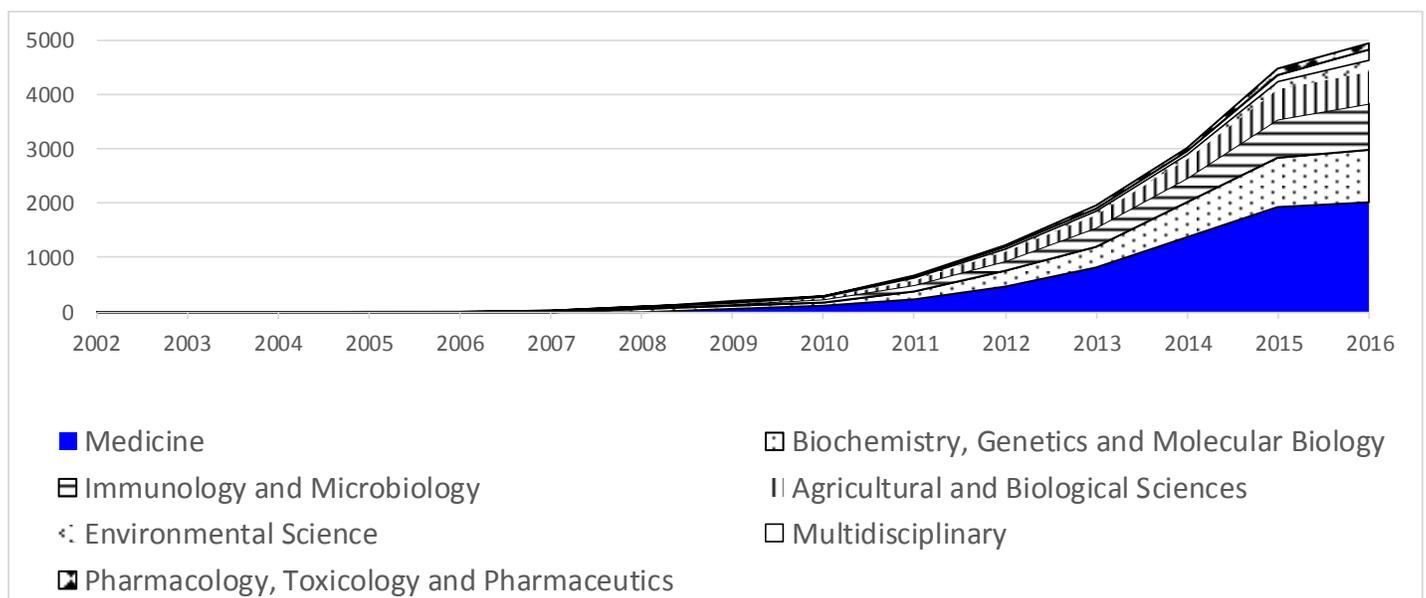

Figure 5. Evolution of microbiome based on driving scientific disciplines

Figure 6 shows the ratio between the number of occurrences concerning microbiome in each discipline in a specific year $t$ and the total number of occurrences of that year (*see* eq. [1]). Considering equation [1] for this research field, the ratio is:

$$weight \ of \ occurrences \ of \ microbiome \ in \ the \ research \ field \ i \ at \ t$$

$$= \frac{occurrence \ of \ microbiome \ in \ the \ research \ field \ i \ at \ time \ t}{total \ of \ number \ of \ occurrences \ in \ microbiome \ at \ t}$$

In particular, figure 6 shows that the research field of human microbiome is originated in medicine, immunology and microbiology in 2002. These disciplines play again a vital role in supporting the evolution of microbiome, but other disciplines over time are supporting the evolutionary growth of the human microbiome, specifically biochemistry, genetics and molecular biology, agricultural and biological sciences. As a matter of fact, 80% of total scientific production in microbiome is due to four research fields: medicine; biochemistry, genetics and molecular biology; immunology and microbiology; and agricultural and biological sciences.

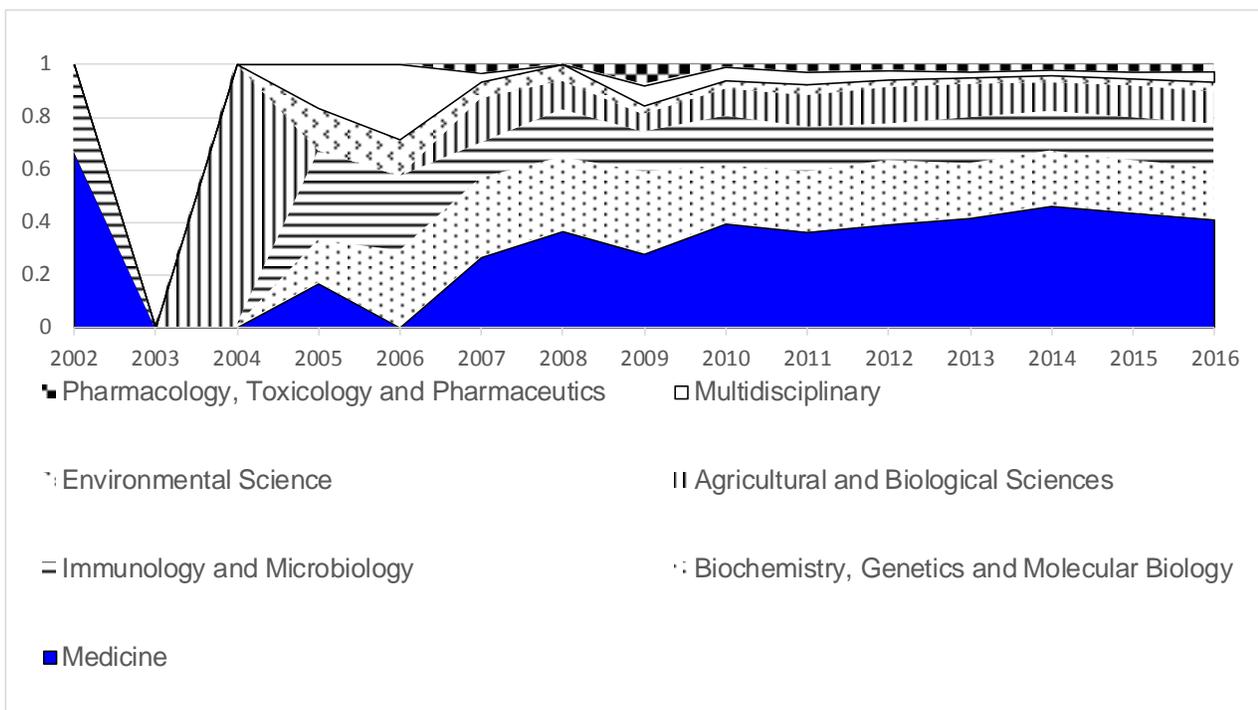

Figure 6. Disciplines supporting the scientific production in human microbiome

This statistical analysis suggests two main findings:

-   Path-dependence of the evolution of this research field from native disciplines in which it is originated.

-   Concentration of the production that supports the evolution of this specific research field in few (four) disciplines.

## 1.2 Evolutionary robotics

Floreano et al. (2008) state that: "Evolutionary Robotics is a method for automatically generating artificial brains and morphologies of autonomous robots. This approach is useful both for investigating the design space of robotic applications and for testing scientific hypotheses of biological mechanisms and processes". The search document of the keyword "evolutionary robotics" in Scopus (2018) shows that the first article concerning this topic is due to American Mathematical Rudolf von Bitter Rucker in 1980 when he worked at mathematical institute of the Ruprecht Karl University of Heidelberg in Germany. Rucker R. in 1980 published an article that can be considered the root source of evolutionary robotics: Rucker, R.B. (1980) "Towards robot consciousness", Speculations in Science and Technology, volume 3, issue 2, June, pages 205-217. This article in 1980 can be considered the origin of this new research field that has grown with accelerated rates during 1990s and 2000s (Figure 7).

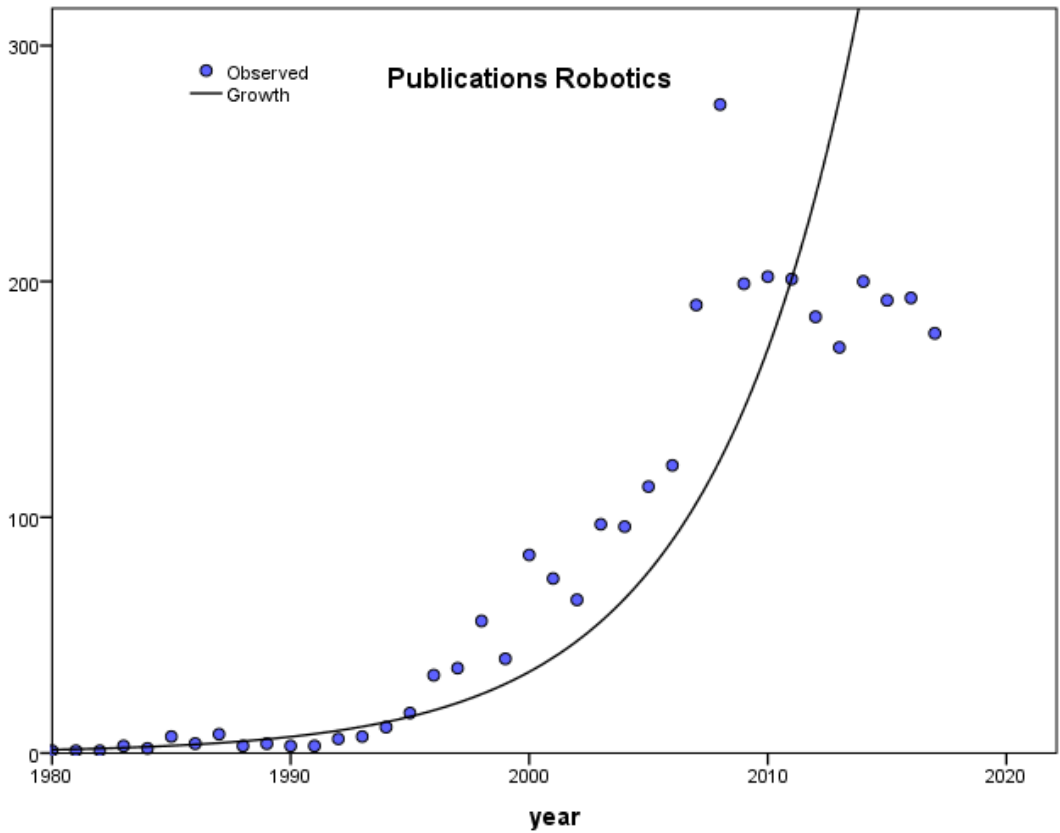

Figure 7. Growth of the research field of evolutionary robotics

The estimated relationship with regression analysis applying a model of growth with equation $y = e^{a+b\,t}$ is in table 2. The $R^2$ value and $t$ ratio are also here very high and model [3] explains more than 95% variance in the data.

Table 2: Estimated relationship of the evolution of the research field "evolutionary robotics"

| Growth model | Constant a (St. Err.) | Coefficient b (St. Err.) | Stand. coefficient | $R^2$ | F (Sign.) |
|---|---|---|---|---|---|
| Dependent variable (D): Annually Publications | | | | | |
| Explanatory variable ○ Time 1980-2017 | −317.31*** (16.99) | 0.16*** (0.009) | 0.953 | 0.91 | 355.73 (0.001) |

Note: ***=Coefficient is significant at p-value<0.001

Figure 8 shows that the trend of evolutionary robotics is driven mainly by computer science, engineering, mathematics, biochemistry, genetics and molecular biology and finally by neuroscience.

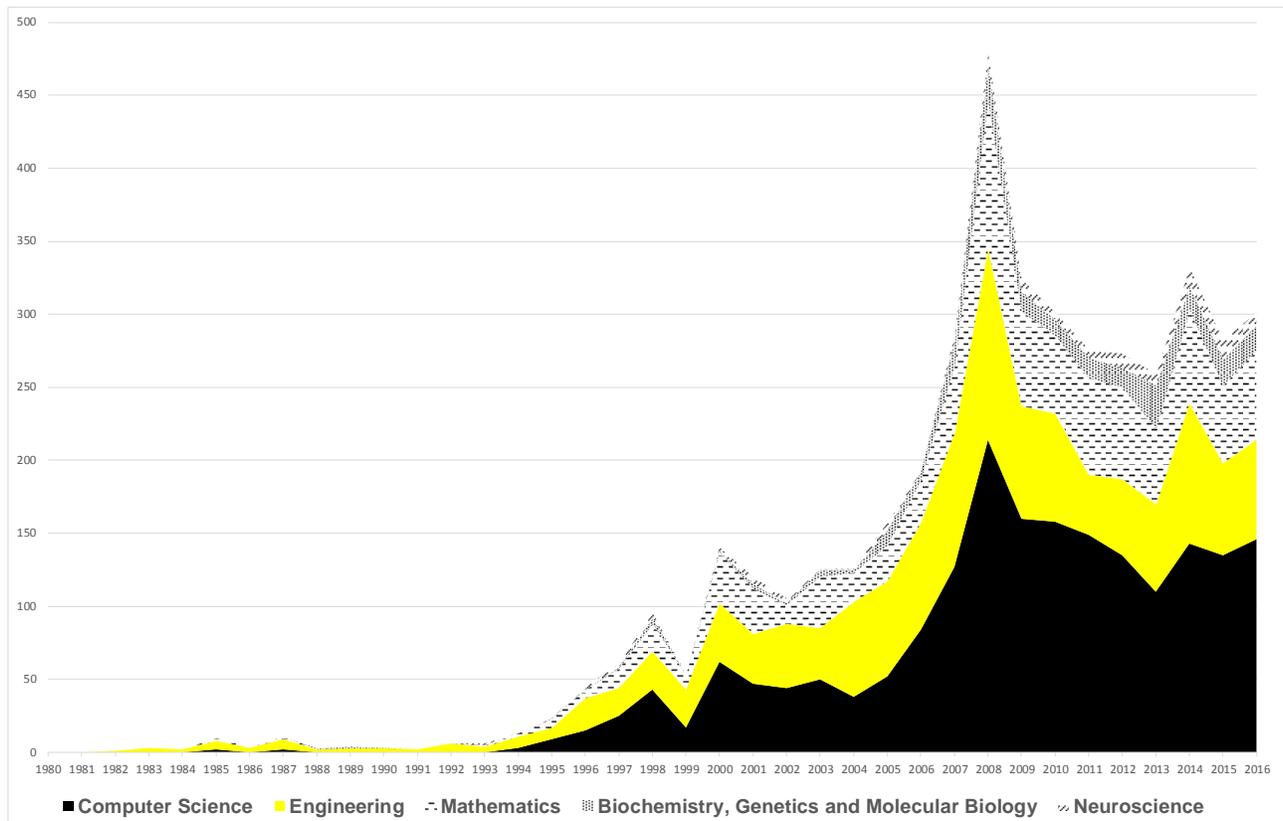

Figure 8. Evolution of evolutionary robotics based on driving disciplines

Figure 9 shows the ratio between the number of occurrences concerning evolutionary robotics in each subject area in a specific year $t$ and the total number of occurrences of that year $t$. In particular, the research field of evolutionary robotics is originated in 1980 within the subject area of mathematics that now has a minor contribution to the evolution of this research field. Subsequently, the research field of engineering has driven the development of this research field of evolutionary robotics, though now it is also reducing its incidence. From 1990s, the computer science, emerged during the social dynamics of electronics and information theory, is playing a critical role in the evolution of this research field, associated with a minor role of mathematics and engineering. In addition, 80% of total scientific production in evolutionary robotics is generated by three research fields: computer science, engineering and mathematics (Figure 9).

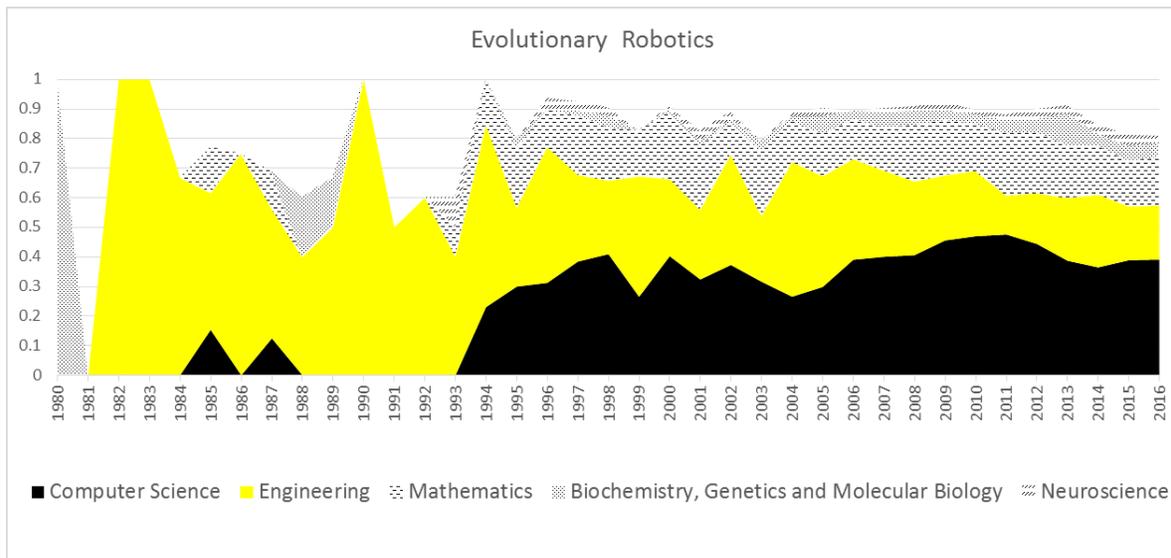

Figure 9. Weight of disciplines in the evolution of the research field of evolutionary robotics

This statistical analysis of evolutionary robotics suggests two main results:

- Decreasing role of native disciplines that have originated the evolutionary robotics (engineering and mathematics), combined with the growing role of computer science from 1990s.

- The production that supports the evolution of this research field is concentrated in three disciplines.

## 1.3 Astrobiology/Exobiology

Astrobiology is a new research field and according to the International Journal of astrobiology (2018) deals with interdisciplinary topics concerning: "cosmic prebiotic chemistry, planetary evolution, the search for planetary systems and habitable zones, extremophile biology and experimental simulation of extraterrestrial environments, Mars as an abode of life, life detection in our solar system and beyond, the search for extraterrestrial intelligence". NASA (2018) argues that the interaction between knowledge acquired from space exploration and astrobiology (then called exobiology) was shown by American molecular biologist Joshua Lederberg that won the Nobel Prize in 1958.

The search document of the keyword "astrobiology" in Scopus (2018) shows that one of the first articles concerning this topic is due to Young R. S. and Johnson J. L from U. S. Army Ballistic Missile Agency (Huntsville, Ala., United States) in 1960. Young R.S. and Johnson J.L (1960) published a letter about "Basic Research Efforts in Astrobiology", in IRE Transactions on Military Electronics, volume MIL-4, issue 2, July 1960, pages 284-287. In 1960, Lenderberg (1960) also published a paper concerning this topic: "Exobiology: Approaches to life beyond the earth" on Science, volume 132, issue 3424, pages 393-400. These articles suggest the origin of this research field in 1960 that is growing more and more (Fig. 10).

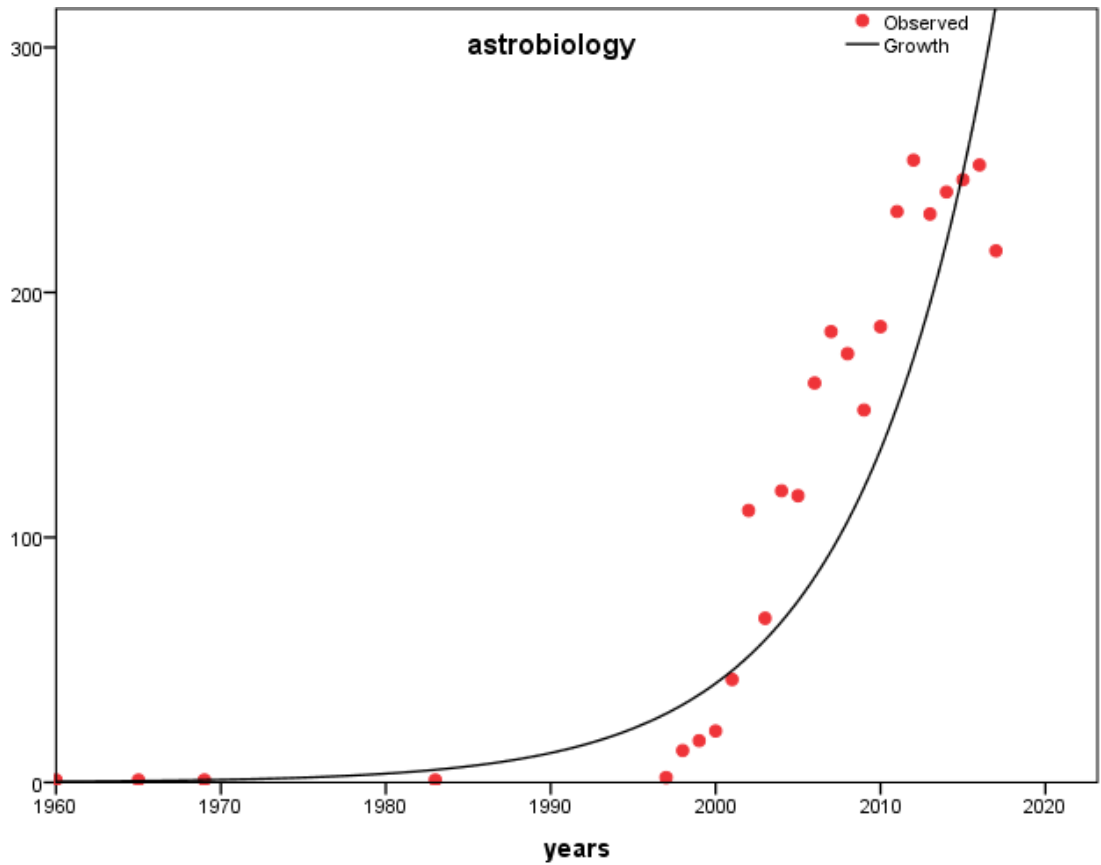

Figure 10. Growth of the research field of astrobiology

The estimated relationship with regression analysis, applying a model of growth with equation $y = e^{a+b\,t}$ is in table 3.

**Table 3**: Estimated relationship of the evolution of the research field "astrobiology"

| Growth model | Constant a (St. Err.) | Coefficient b (St. Err.) | Stand. coefficient | $R^2$ | F (Sign.) |
|---|---|---|---|---|---|
| Dependent variable (D): Annually Publications | | | | | |
| *Explanatory variable* ○ Time 1960-2017 | −238.81*** (21.85) | 0.12*** (0.011) | 0.918 | 0.84 | 123.27 (0.001) |

*Note*: ***=Coefficient is significant at *p-value<0.001*

The $R^2$ value and *t* ratio here are also very high and model [3] explains more than 91% variance in the data.

Figure 11 shows that the trend of astrobiology is driven mainly by earth and planetary sciences, physics and astronomy, agricultural and biological sciences, finally engineering.

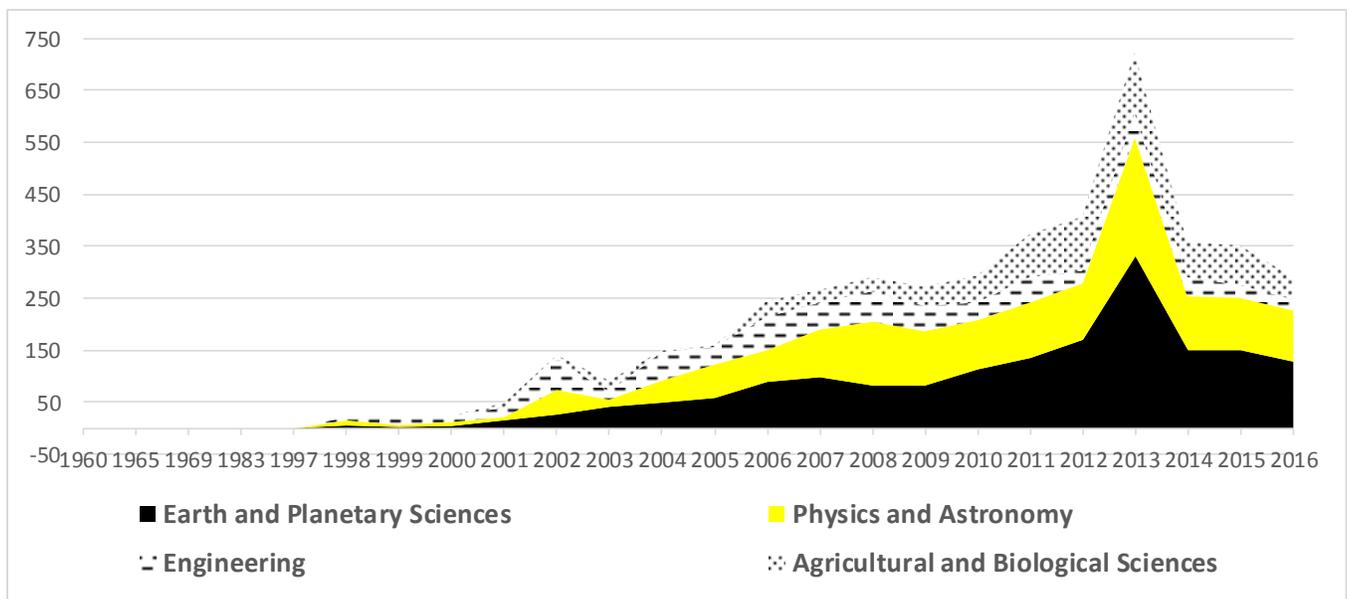

Figure 11. Evolution of astrobiology based on disciplines

Figure 12 shows the ratio between the number of occurrences concerning astrobiology in each discipline in a specific year $t$ and the total number of occurrences at year $t$. In particular, astrobiology is originated in 1960 mainly in the research field of engineering that now has a minor contribution to the evolution of this research field. Subsequently, the disciplines of earth and planetary sciences, physics and astronomy, associated with agricultural and biological sciences are driving the development of astrobiology. Moreover, 80% of total scientific production in astrobiology is generated by five scientific disciplines (Fig. 12).

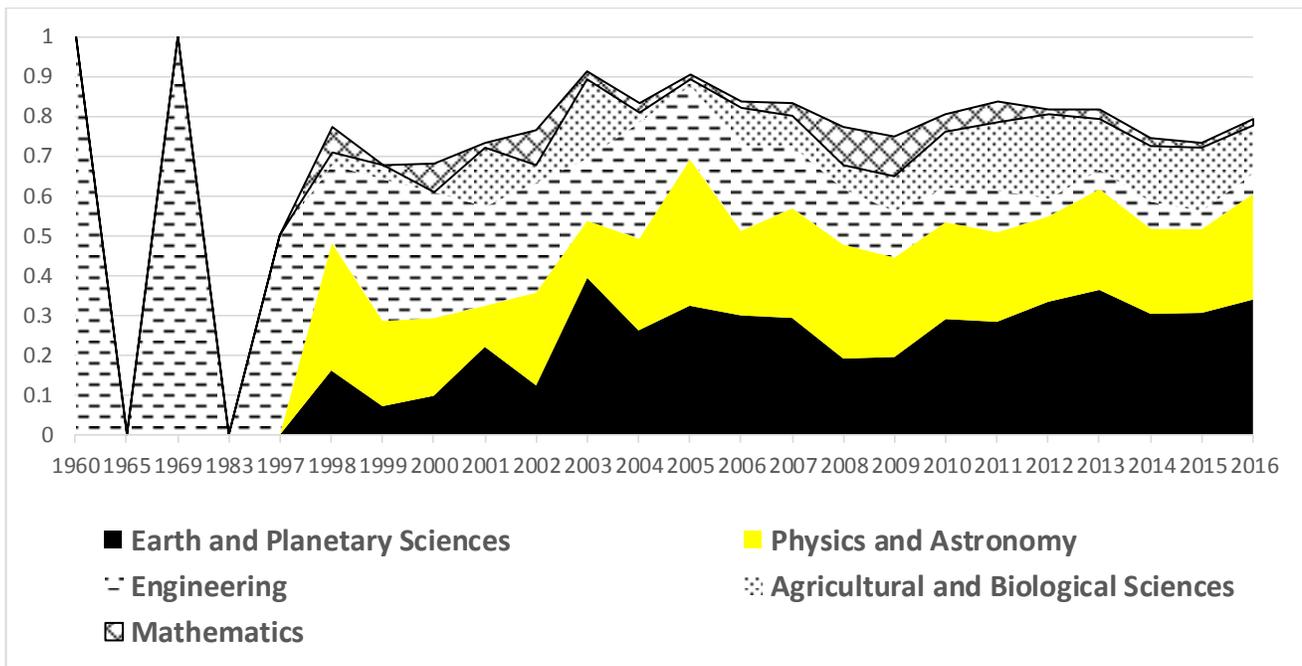

Figure 12. Disciplines supporting more than 80% of scientific production in astrobiology

Exobiology is a similar research field to astrobiology and the concept is used interchangeably. Scatter of data here is more spread over time (Fig. 13).

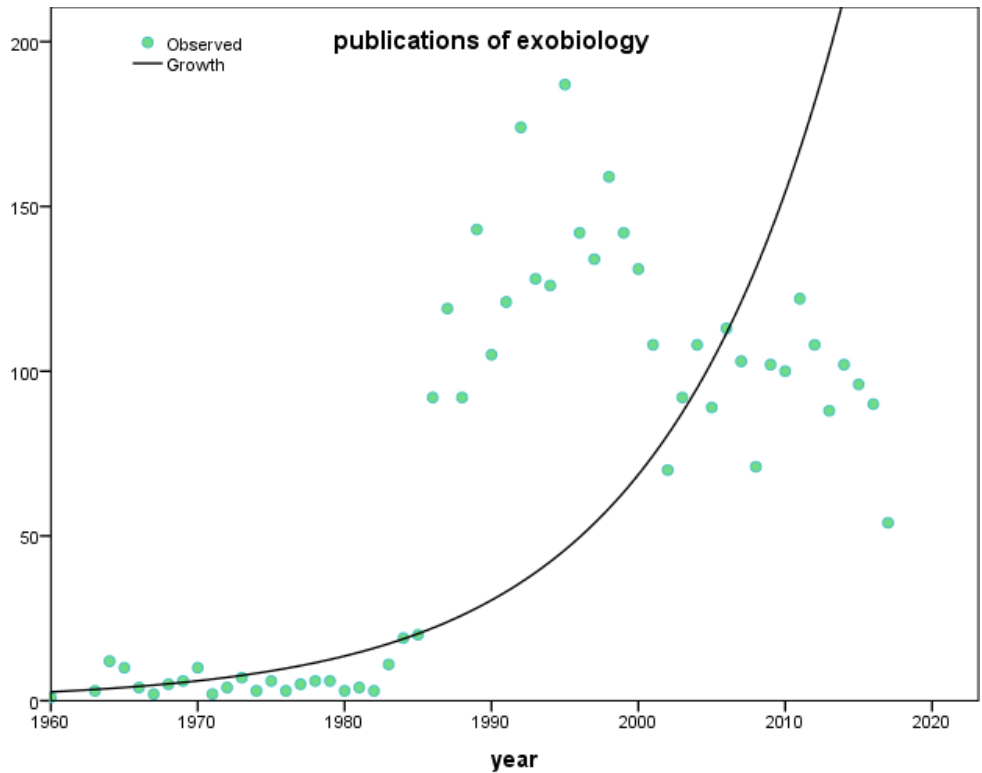

Figure 13. Growth of the research field of exobiology

The estimated relationship with regression analysis applying a model of growth [3] with equation $y = e^{a+b\,t}$ is in table 4.

**Table 4**: Estimated relationship of the evolution of the research field "exobiology"

| Growth model | Constant a (St. Err.) | Coefficient b (St. Err.) | Stand. coefficient | $R^2$ | F (Sign.) |
|---|---|---|---|---|---|
| Dependent variable (D): Annually Publications | | | | | |
| Explanatory variable ○ Time 1960-2017 | −157.86*** (15.31) | 0.081*** (0.008) | 0.82 | 0.67 | 110.90 (0.001) |

*Note*: ***=Coefficient is significant at *p-value<0.001*

The $R^2$ value and *t* ratio here are also high and model [3] explains about 80% variance in the data.

Figure 14 shows the ratio between the number of occurrences concerning exobiology in each discipline in a specific year *t* and the total number of occurrences at year *t*. In particular, the evolution of exobiology is driven by multidisciplinary. The most important disciplines here are earth and planetary sciences, agricultural and biological sciences, whereas physics and astronomy, biochemistry, genetics and molecular biology are decreasing over time. Medicine is also a discipline supporting the evolution of exobiology, though at lesser extent than other ones. More than 80% of scientific production in exobiology is driven by about six disciplines.

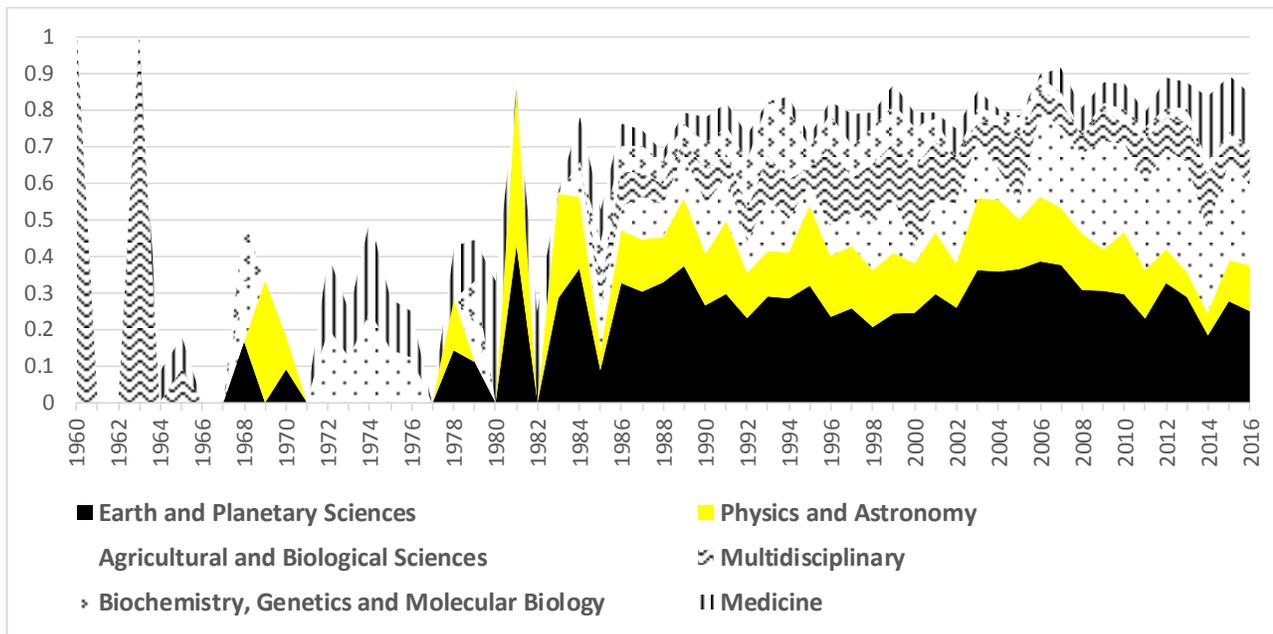

Figure 14. Disciplines supporting more than 80% of scientific production in exobiology

This statistical analysis of the evolution of astrobiology suggests two main findings:

- Decreasing role of native disciplines that have originated the astrobiology (e.g., engineering), combined with the growing role of earth and planetary sciences and agricultural and biological sciences from 1990s and steady-state contribution of physics and astronomy over time.

- The evolution of astrobiology is concentrated in five research fields.

Exobiology has a similar pattern of growth. It is driven by earth and planetary sciences, agricultural and biological sciences, and medicine. In addition, the research field of exobiology seems to be more interdisciplinary because 80% of the production is due to six disciplines (Fig. 14).

**Empirical laws of the evolution of research fields**

This inductive study here seems to support the following empirical laws of the evolution of scientific fields.

[1]  *The first law* states that the evolution of a specific research fields is driven by few disciplines (3-5) that generate more than 80% of documents (concentration of the scientific production)

[2]  *The second law* states that the evolution of research fields is path-dependent of a critical discipline (it can be a native discipline that has originated the research field or a new discipline emerged during the social dynamics of science).

[3]  *The third law* states that a research field can be driven during its evolution by a new discipline originated by a process of specialization of a specific science.

*The causes and consequences of empirical laws: concentration, scientific guideposts and interdisciplinary as engines of the evolution of research fields*

The production of research fields is generally concentrated in 3-5 disciplines. In fact, case study of research fields here shows that more than 80 % of the production is driven by few disciplines that support the evolution over time. The distal cause of this concentration can be due to leading scientists and scientific communities. In fact, Levin and Stephan (1991) argue that scientific productivity is asymmetrically distributed throughout the population of researchers (Allison and Stewart, 1974; David, 1994; Fox, 1983). A study by Ramsden (1994) about Australian universities shows that, over a 5-year period, 14% of the total number of researchers produced 50% of the publications, while 40% of researchers produced 80% of publications. A determinant of the concentration within specific disciplines is due to characteristic of the winner-take-all that generates inequality in scientific production. In fact, scientific research has extreme inequality with regard to productivity because high productivity of some researchers generates cumulative learning processes (Matthew effect in science, Merton, 1957, 1968). This effect shows how researchers in specific disciplines that accomplish prominent results have an initial advantage over others and increased chances of obtaining further financial support as well as of accomplishing further discoveries.

In addition, results suggest that critical disciplines are the driving force of specific research fields providing scientific guideposts and avenue that lay out certain definite paths of development. In general, chance indicates which scientific guideposts are chosen in the course of social dynamics of science. In other words, development of research fields seems to be governed by a common system of evolution. Normally, evolution within any given scientific field leads to the creation of a certain research programme, *sensu* Lakatos (1978). The research programme guides the subsequent steps in the process of scientific development. Hence, discoveries are due to incremental development of a basic pattern of research programme (scientific guidepost). Vital corollary to above propositions is that science advances are due to systematic R&D activity, rather a random method, which generates distinct pathways of the evolution of research fields. The scientific guideposts point to scientific avenue of research fields. This dynamics suggests that the evolution of research fields can be supported by increasing the investment in specific research programmes in order to produce science advances and scientific breakthroughs to maximize benefits in society.

The evolution of science also induces the emergence of new disciplines by either from one specific discipline or through the combination of multiple disciplines (cf. also, Jamali and Nicholas, 2010; Riesch, 2014). US National Research Council (2014) states that interdisciplinarity is a key process to spur breakthroughs by research teams in both theoretical and applied sciences. In fact, some new disciplines have been established with an intrinsically interdisciplinary nature, such as nanoscience, nanotechnology, biotechnology, cognitive science, computational biology, etc. (cf. Jeffrey, 2003; Van Raan, 2000, 2000a). Battard (2012) argues that emerging research fields, such as nanotechnology, involve several disciplines around the same complex problem, as well as "laboratories are technological hubs through which scientists converge from multiple scientific backgrounds" (Battard, 2012, p. 235). In addition, traditional disciplines, such as chemistry, physics and biology have been shown to be highly interdisciplinary as well (Boyack et al., 2005). The characteristic of interdisciplinarity in both new and traditional scientific fields, in the light of "big science" (De Solla Price, 1986) challenge, supports emerging research fields for the investigation of complex problems necessary to accelerate problem solving and benefits in the modern societies (cf., Tijssen, 2010).

## Discussion and concluding observations

Science is a complex and stratified progress that changes structure and specific weight within society depending on historical period. Since Galileo (1564-1642), Kepler (1571-1630) and Newton (1642-1727) an unparalleled proliferation of science advances is generated as never before in history and it is no coincidence that this happened in conjunction with major socioeconomic events and a new economic view of progress (Coccia and Bellitto, 2018; Woods, 1907; Seligman, 1902). Starting from this, the evolution of science was a cause but also a consequence of the economic vector. The evolution of scientific fields in physics, economics, biology, etc. has a vital role to explain the overall progress of science in society. Since 1960s, many studies describe, explain and predict different characteristics of science (cf., Börner and Scharnhorst, 2009; Börner et al., 2011; Freedman, 1960; Scharnhorst, et al., 2012, Sun et al., 2013;). A vital question in social studies of science is what is meant by progress in science? During the Twentieth century, the role of science in society has grown so much that it has become functional to civil and military state institutions, as well as to world production, technological and economic processes (cf., Coccia and Wang, 2016; Stephan, 1996). The evolution of science and research fields can be explained by using the theoretical framework of the Gestalt psychology given by four elements (see Basalla, 1988, p. 23): 1) Perception of the problem: an incomplete pattern in need of resolution is recognized; 2) Setting stage: data related to the problem is assembled; 3) Act of insight: a mental act finds a solution to the problem; 4) Critical revision: overall exploration and revision of the problem and improvements by means of new acts of insight. In general, the mental acts of insight solve problems and support the evolution of science that is the results of a cumulative change (Popper, 1959). Moreover, science, in the pragmatic approach of the philosophy of science is conceived in terms of its aims. In this context the evolution of science is due to goals that are presumably attainable objectives of science and ideals, which are unattainable but approachable within any limits. The ideals and sub-ideals represent steps in the pattern of science's progress that must be approached as science itself progresses (West Churchman and Ackoff, 1950).

In response to the first question stated in the introduction, how do research fields evolve?

It seems that the origin of scientific fields is due to both endogenous processes in science driven by specific disciplines, interaction between disciplines, convergence of disciplines, scientific discoveries and breakthroughs, and exogenous factors given by socioeconomic elements (cf., Coccia, 2015, 2017).

With respect to the second question (What drives the evolution of a research field? One or more disciplines? The same discipline or different disciplines over time?), the answer, based on empirical evidence here, also seems clear: the driving forces of research fields are few disciplines, along a pathway affected by path-dependence of native and associated disciplines. In fact, the evolution of scientific fields can be guided by native disciplines and/or new ones that emerge from a specialization process of science and/or merger between different branches of sciences. The evolution of research fields is a part of the natural process of the social dynamics of science in society (Sun et al., 2013) guided mainly by social interactions between scientists in an international network of collaboration. As a matter of fact, scientific research is becoming more international, more interdisciplinary, more conducted by groups within and between networks of researchers (Adams, 2012, 2013; Coccia and Wang, 2016; Gibbons *et al.,* 1994; Newman, 2001, 2004; Pan et al., 2012).

The results of the analysis here about the evolution of scientific fields are:

1. The evolution of research fields is driven by few disciplines that concentrate more than 80% of scientific production.

2. Path-dependence of the evolution of specific research field from native disciplines in which it is originated and/or new discipline.

3. In some research fields, the decreasing role of native disciplines that have originated them is balanced by the vital and increasing role of new disciplines emerged during the social dynamics of applied and basic sciences that support the evolutionary growth of emerging research fields.

These results are systematized in three empirical laws that can explain and generalize the long-run behavior of the evolution of scientific fields. Consequences of these findings are that research fields attempt to emerge from the specialization of disciplines and fusion of social communities of scholars within an international collaboration network. These factors feed social interactions between scholars of different disciplines in shaping the overall dynamics of science.

Science is indeed a multi-layered system involving communities of scientists engaged in international research collaboration in order to produce new knowledge and/or science advances within and between scientific fields (Coccia and Wang, 2016; Lee and Bozeman, 2005). In general, evolution of scientific fields can be originated from convergence between applied and theoretical sciences (Coccia and Wang, 2016), paradigm shifts (Khun, 1962), research programmes (Lakatos, 1978), branching due to scientific breakthroughs, new discoveries and technologies, fractionalization and specialization of general disciplines (De Solla Price, 1986; Mulkay, 1975; Dogan and Pahre, 1990; van Raan, 2000). Some theoretical frameworks argue that new social groups of scientists can be the driving force of the evolution of emerging scientific fields (Crane, 1972; Guimera et al., 2005; Wagner, 2008).

However, the question of progress in science is always associated with the role that science itself plays within contemporary society. Lyotard (1979) argues that the state is willing to spend a lot of money in R&D investments to make science a driver of economic growth in society. In this context the birth, growth and decline of research fields play a vital role in social and economic change of nations. As a matter of fact, the findings suggested by the evidence here are important to explain the social dynamics of science for allocating economic resources directed to accelerate science and technological advances that amply benefits in society. Callon (1994) argues that public subsidy to support emerging research fields is needed, though results can be uncertain and/or achieved only in the long run, such as in gravitational astronomy that studies the sources of the universe. The understanding of the evolution of scientific fields can also support economic growth and competitive advantage of nations (Coccia, 2017b, 2017c, 2018, 2018b). In fact, investments in path-breaking fields of research, and associated education, are a vital political economy of R&D to increase productivity both of firms and of national innovation systems (Gibbons et al., 1994). In addition, in this context, research policies of leading nations should be directed to support the evolution of new research fields with a mission-oriented approach (Ergas, 1987) to achieve, shift and sustain the frontiers of science and knowledge (cf., Coccia, 2015; 2017).

Overall, then, this study endeavors to explain properties of the evolution of scientific fields to predict the long-run behavior directed to support a research policy for enhancing the positive impact of science in society. In fact, the evolution of science is associated with techno-economic processes directed to readjustment of human institutions to new economic and social bases (cf., Woods, 1907, pp. 810-811; Bernstein, 1893). In general, the underlying factors of science progress and evolution of research fields are human wants and human control of nature (cf., Woods, 1907). Evolution of science should be the forerunners of a full realization of the meaning and possibilities of life of individuals in society. This realization of the life of individual is achieved in appropriate social contexts with education, culture and sustainable environment (Coccia and Bellitto, 2018).

To conclude, it would be elusive to limit the evolution of scientific fields to endogenous factors of the dynamics of science. The birth, growth and decline of research fields are due to manifold factors represented by social context, economic growth, power of nations, military and political tensions to prove scientific and technological superiority in a context of social and economic change affecting human behavior (cf., Small, 1905, p. 682). Sun et al. (2013, p. 3) show: "the correspondence between the social dynamics of scholar communities and the evolution of scientific disciplines". As a matter of fact, the evolution of scientific fields is due to the expanding content of the human life-interests whose increasing realization constitutes progress that characterizes the human nature from millennia (Woods, 1907, pp. 813-815). Hence, the whole process of science progress is driven by the increasingly effective struggle of the human mind in its efforts to raise superior to the exigencies of the external world and attitude to satisfy human desires, solve problems and achieve/sustain power in society. The results here provide an empirical evidence of basic properties of the evolution of scientific fields and in general of the dynamics of science. However, these conclusions are of course tentative because we know that other things are not equal over time and space. One of the main problems is the difficulty of formally defining the notion of scientific field that is a concept that can change over time during the social dynamics of science. Therefore, identifying general patterns of science and scientific fields, affected by manifold and complex factors, at the intersection of economic, social, psychological, anthropological, philosophical, religious, and perhaps biological characteristics of human being, is a non-trivial exercise. There is need for much more detailed research into the relations between science, technology and society to explain and predict the evolution of research fields that is more and more important for human progress.

Coccia M. (2018) The laws of the evolution of research fields